\documentclass[prb,twocolumn,preprintnumbers,amsmath,amssymb,superscriptaddress]{revtex4}

\usepackage{graphicx}
\usepackage{dcolumn}
\usepackage{bm}

\begin{document}

\title{Pinned Low Energy Electronic Excitation in  Metal Exchanged Vanadium Oxide Nanoscrolls}

\author{J. Cao}
\affiliation{Department of Chemistry, University of Tennessee,
Knoxville, Tennessee 37996-1600}
\author{J. L. Musfeldt}
\email{musfeldt@utk.edu} \affiliation{Department of Chemistry,
University of Tennessee, Knoxville, Tennessee 37996-1600}
\author{S. Mazumdar}
\affiliation{Department of Physics, University of Arizona, Tucson,
Arizona 85721}
\author{N. Chernova}
\address{Department of Chemistry and Institute for Materials Research, State University of New
York, Binghamton, New York 13902-6000}
\author{M. S. Whittingham}
\address{Department of Chemistry and Institute for Materials Research, State University of New
York, Binghamton, New York 13902-6000}

\date{\today}

%
%
\begin{abstract}

We measured the optical properties of mixed valent vanadium oxide
nanoscrolls and their metal exchanged derivatives in order to
investigate the charge dynamics in these compounds. In contrast to
the prediction of a metallic state for the metal exchanged
derivatives within a rigid band model, we find that the injected
charges in Mn$^{2+}$ exchanged vanadium oxide nanoscrolls are
pinned. A low-energy electronic excitation associated with the
pinned carriers appears in the far infrared and persists at low
temperature, suggesting that the nanoscrolls are weak metals in
their bulk form, dominated by inhomogeneous charge
disproportionation and Madelung energy effects.

\end{abstract}
\narrowtext \maketitle
\clearpage


The discovery that low-dimensional inorganic solids can curve or
fold into nanoscale objects provides an exciting opportunity to
investigate bulk versus nanoscale chemistry using molecular-level
strain as the tuning parameter. \cite{Rao2003, Halford2005,
Tenne2006} These materials are targets of intensive research
efforts, driven by the need to further miniaturize electronic
devices and the potential to exploit unusual mechanical and optical
properties. Some of the beautiful, flexible, and functional
nanomorphologies include tubes, wires, octahedra, particles,
urchins, and spheres. \cite{Rao2003, Halford2005, Tenne2006,
Moses2003, Wu2001, Krumeich1999, Muhr2000, Worle2002, Dwyer2006}
They offer molecular-level control of size, shape, mechanical
response, and chemical composition as well as unusual confinement
effects due to finite length scales.

\begin{figure}
\includegraphics[width=3.6 in]{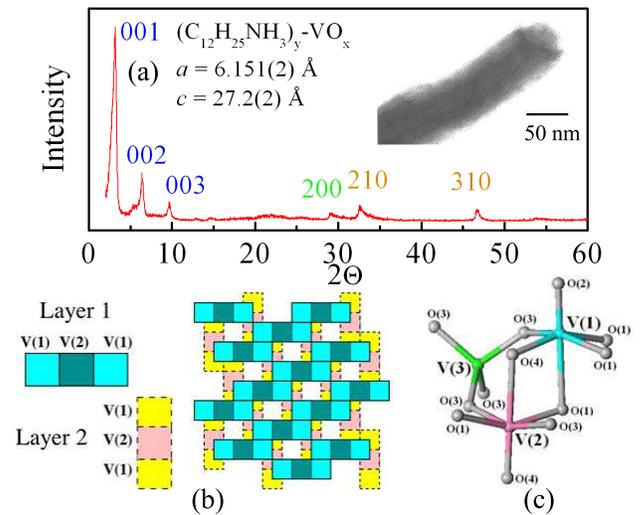}
\caption{\label{fig_dvostruc}(color online) (a) Representative x-ray
powder diffraction scan of (C$_{12}$H$_{25}$NH$_3$)$_y$-VO$_x$
scrolls showing only 00l and weak hk0 reflections. The inset
displays a representative TEM image with lattice fringing and a
center cavity. (b) Probable double-layer structure of the VO$_x$
scrolls, analogous to the BaV$_7$O$_{16}$ model compound.
\cite{Wang1998} (c) Close-up view of the local structure around the
V centers, showing the arrangement of octahedra and tetrahedra in
the BaV$_7$O$_{16}$ model compound. \cite{Wang1998}}
\end{figure}

Among the transition metal oxides, vanadates show particularly rich
chemistry due to the tunable vanadium oxidation state and flexible
coordination environment, which ranges from octahedral to square
pyramidal to tetrahedral with increasing vanadium oxidation state.
\cite{Zavalij1999} Vanadium oxides form many layered and nanoscale
compounds with open structural frameworks, making them prospective
materials for
ion intercalation, exchange and storage. \cite{Muhr2000, Worle2002,
Dwyer2006, Wang2006} The nanoscale vanadates of interest here,
(amine)$_y$VO$_x$, are formally mixed-valent\cite{Krumeich1999} with
$x$ $\sim$ 2.4 and $y$
$\sim$ 0.28.  Stoichiometric considerations do not, of course,
distinguish between a true mixed-valent
V$^{4.5+}$ state that may be metallic and a state with an
inhomogenous charge distribution, consisting of V-ions with distinct
small and large charges. Such charge disproportionation is common in
oxides of Ti and V with formal oxidation states of Ti$^{3.5+}$ and
V$^{4.5+}$,  respectively, and the convention is to assign integer
charges (Ti$^{3+}$ and Ti$^{4+}$; V$^{4+}$ and V$^{5+}$) to the ions
with different oxidation states.\cite{Marezio73, Chakraverty78,
Ohama97} We adopt the same convention here, and consider the
possibility of charge disproportionation in the nanoscale vanadates.

The compounds we consider are actually scrolls, consisting of
vanadate layers between which organic molecules are intercalated.
\cite{Krumeich1999, Muhr2000, Worle2002, Niederberger2000,
Reinoso2000} The size of the amine or diamine template determines
scroll winding, providing an opportunity to tune the size of these
materials. The typical scrolls are $\sim$15-100 nm in diameter,
containing up to 30 vanadium oxide layers (inset of
Fig.~\ref{fig_dvostruc}(a)). The exact crystal structure is
complicated and is not completely understood.
Figure~\ref{fig_dvostruc}(a) shows a representative x-ray powder
diffraction pattern for the (C$_{12}$H$_{25}$NH$_3$)$_y$-VO$_x$
scrolls. The (00l) reflections from the vanadium oxide layers are
clear, but the (hk0) reflections from the atomic structure within
each layer are weak.
The data
suggest a tetragonal lattice of basal repeat size 6.144 \AA, closely
related to that of BaV$_7$O$_{16}$, which has a planar structure of
basal size 6.160 \AA. \cite{Wang1998} The arrangement in
BaV$_7$O$_{16}$ is shown in Fig.~\ref{fig_dvostruc}(b).
\cite{Wang1998} It contains zig-zag chains of edge-sharing VO$_5$
square pyramids. These chains share some corners with each other to
form a two dimensional sheet. The sheets also share corners with
VO$_4$ tetrahedra, bringing the two VO$_x$ layers together to form a
characteristic double sheet. Figure~\ref{fig_dvostruc}(c) displays
the local structure around the V centers in the BaV$_7$O$_{16}$
model compound. Each two-dimensional layer contains two octahedrally
coordinated vanadium atoms, V(1) and V(2), with one tetrahedrally
coordinated V(3) occurs in between the layers.

Scrolled vanadates exhibit very interesting physical properties
including a large spin gap, diameter-dependent optical features, and
potential battery and optical limiting applications. \cite{Cao2004,
Krusin-Elbaum2004, Nordlinder2006, Xu2002, Vavilova2006} Recently,
electron- and hole-doped vanadium oxide nanoscrolls were reported to
be 300 K ferromagnets, raising fundamental questions about
the underlying magnetic exchange mechanism. \cite{Krusin-Elbaum2004}
It was further suggested that charge-injection via ``doping''
introduces free carriers into the pristine Mott insulating scrolls,
resulting in partially filled bands. The ``doped'' nanoscrolls are
predicted to exhibit Drude-type metallic behavior within this
picture. \cite{Krusin-Elbaum2004} Direct measurement of the
low-energy electronic structure is clearly important to test this
prediction.
Further, the rigid band model raises fundamental questions about the
chemical nature of the ``doping'' or ion exchange process in the
scrolled vanadates. \cite{Reinoso2000, Lutta2002, Krusin-Elbaum2004,
Filho2004}
From the chemical perspective, metal intercalation does not appear
as simple as ``putting in electrons''; both reduction of vanadium as
well as ion exchange with the proton on the amine are possible
consequences of the exchange process. To date, only Mn$^{2+}$ has
been shown to completely replace the organic template, even though
some other ions such as Na$^+$, Li$^+$, Zn$^{2+}$, Cu$^{2+}$, and
Ca$^{2+}$  can partially replace the amine template.
\cite{Reinoso2000, Lutta2002, Filho2004}

In order to understand the fundamental charge dynamics and 
test the applicability of the rigid band model in vanadium oxide
nanoscrolls, we investigated the variable temperature optical
spectra of the pristine and Mn$^{2+}$ exchanged materials. In
contrast to the rigid band model expectation for a metallic state,
we find that charge is pinned in the metal substituted scrolls. The
low-energy electronic excitation associated with the pinned carriers
appears in the far infrared and persists at low temperature,
suggesting that the nanoscrolls are weak metals in their bulk form,
dominated by inhomogeneous charge disproportionation and Madelung
energy effects. We propose an alternate
model for charge injection to account for our observations. Analysis
of the vibrational properties shows that the 575 cm$^{-1}$ V-O-V
equatorial stretching mode is very sensitive to metal substitution,
indicating that ion exchange modifies both the local curvature and
charge environment. Charge effects also shift excitations in the
higher energy optical conductivity.


Vanadate nanoscrolls were prepared by an initial sol-gel reaction
followed by hydrothermal treatment, as described previously.
\cite{Lutta2002, Cao2004} Appropriate amines (C$_n$H$_{2n+1}$NH$_2$
with $n$=4 - 18) were employed as the structure-directing agent. Ion
exchange was preformed by stirring a mixture of vanadate nanoscrolls
and MCl$_2$ (M=Mn$^{2+}$, Zn$^{2+}$, and Na$^+$) for 2 hours in
ethanol/water mixture. The spectroscopic experiments were carried
out over a wide energy range (4 meV - 6.2 eV; 30 - 50000 cm$^{-1}$)
using a series of spectrometers. \cite{Cao2004} Variable temperature
work was done between 4.2 and 300 K using an open-flow helium
cryostat and temperature controller. The optical constants were
calculated by a Kramers-Kronig analysis of the measured reflectance:
${\tilde{\epsilon}}(\omega) =
{\epsilon}_1(\omega)+i{\epsilon_2}(\omega)={\epsilon}_1(\omega)+\frac{4\pi
i}{\omega}{\sigma}_1(\omega)$. \cite{Wooten1972}


\begin{figure}
\includegraphics[width=3.9 in]{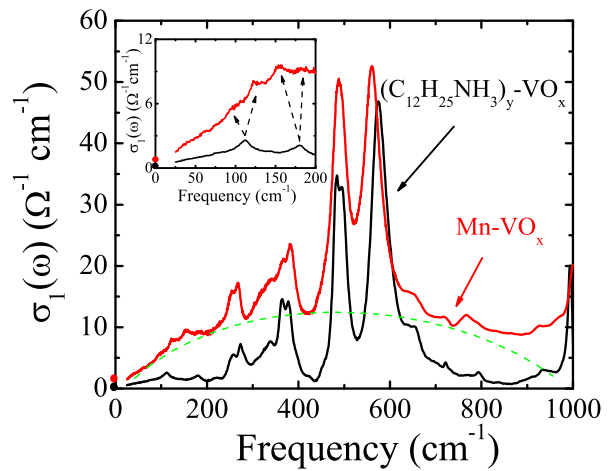}
\caption{\label{fig_dvolowfre}(color online) 300 K optical
conductivity of pristine VO$_x$ scrolls and the Mn$^{2+}$ exchanged
compound in the far-infrared regime. Mode assignments include: V-O
axial stretching at 945 and 995 cm$^{-1}$, weak V-O-V equitorial
stretching at 790 and 866 cm$^{-1}$,   V-O-V equitorial stretching
at 585, 652, and 727 cm$^{-1}$,  V-O-V axial stretching at 486 and
498 cm$^{-1}$, V-O bending at between  179 and 381 cm$^{-1}$, 
and screw-like motion of the scroll at 113 cm$^{-1}$.\cite{Cao2004}
The dashed green line guides the eye, highlighting the additional
localized carrier contribution in the substituted scrolls. The inset
shows a magnified view of the low-frequency response of pristine and
Mn$^{2+}$ exchanged scrolls. The extrapolated dc conductivity is
indicated by filled circles at $\omega$ = 0 in both panels.}
\end{figure}

Figure~\ref{fig_dvolowfre} shows the optical conductivity of the
pristine VO$_x$ scrolls and the Mn$^{2+}$ substituted compounds at
room temperature. The pristine scrolls display semiconducting
behavior, with clearly resolved vibrational modes below 1000
cm$^{-1}$ and a low background conductivity. These modes were
previously assigned as V-O-V axial and equatorial stretching, V-O
bending, and screw-like motion of the scrolls. \cite{Cao2004}
 Subtle spectral changes occur when
the amine template is exchanged for Mn$^{2+}$, Zn$^{2+}$,  or
Na$^+$. 
In the Mn$^{2+}$ substituted compound (Fig.~\ref{fig_dvolowfre}),
the optical conductivity develops a broad electronic background
in the far infrared region, with strong phonons riding top of this
excitation. The Fano-like lineshapes of several phonons, especially
those near 250 and 380 cm$^{-1}$, indicate that the background is
electronic in nature.\cite{Fano1961}  This  localized electronic
excitation has a width of $\sim$400 - 500 cm$^{-1}$.
Extrapolating $\sigma_1(\omega)$ to zero frequency, we find
$\sigma_1(0)$ $\sim$1 $\Omega^{-1}$ cm$^{-1}$. Such a pinned carrier
response is characteristic of a weak or ``bad'' metal rather than a
traditional Drude metal and has been observed in other complex
oxides.\cite{Liu1999, Choi2004,  Zhu2002, Cao2006}
 We obtain similar
results at low temperature, suggesting that the vanadium centers in
both pristine and ion-exchanged scrolls are charge
disproportionated, with V$^{4+}$ and V$^{5+}$ centers rather than
the mixed valent state (V$^{4.5+}$) that might be anticipated for a
highly conducting material.

Based on the observation of a pinned electronic excitation in the
far infrared, the ion exchange processes does add carriers to the
scrolled vanadates. The carriers are not, however, mobile in the
ion substituted scrolls. 
Single scroll scanning tunneling microscope measurements are in
progress to complement this spectroscopic work with local dI/dV
measurements. Previous individual tube measurements of Li$^+$
exchanged scrolls actually showed a decrease in conductivity with
Li$^+$ hole doping.\cite{Krusin-Elbaum2004}

\begin{figure}[t]
\includegraphics[width=3.4 in]{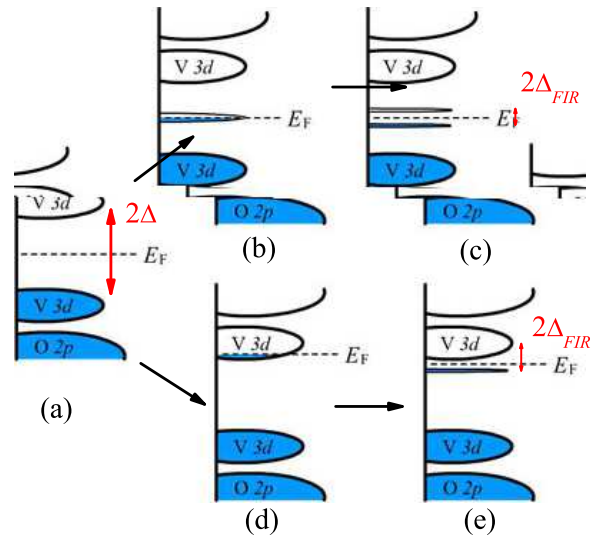}
\caption{\label{fig_dvomodel}(color online) Schematic representation
of possible electronic structure changes
resulting from the ion exchange process in the scrolled vanadates.
(a) Band structure of pristine vanadium oxide nanoscrolls. Here,
$2\Delta$ denotes the $\sim$0.5 eV optical gap.\cite{Cao2004} (b)
Schematic view in which ion exchange adds carriers to a new
defect-level band that forms near $E_F$. (c) Cartoon view in which
the new charge defect band splits due to the Madelung energy
difference that exists in any charged system, electron-electron
interactions, and/or chemical disorder effects. Here,
$2\Delta_{FIR}$ represents the pinned low-energy electronic
excitation. This structure will likely have a relatively narrow
bandwidth because the excitation is between split defect levels. (d)
Alternate approach to the exchange process in which ion exchange
adds carriers within the rigid band model.\cite{Krusin-Elbaum2004} A
metallic state might be anticipated to arise from the
partially-filled band. (e) Schematic view in which the V $3d$ band
splits
due to  chemical disorder effects. Here, $2\Delta_{FIR}$ represents
a possible low energy excitation, although it might be anticipated
to display a wider band width than in scheme (c) due to joint
density of states considerations. }
\end{figure}

We now discuss the evolution of the electronic structure of the
nanoscrolls
due to the metal exchange process within a heuristic
correlated-electron model.
%
We use an ``effective'' band picture, even though a localized
description may be more suitable for charge-disproportionated
V-oxides, \cite{theory} in order to maintain continuity with the
existing literature. \cite{Krusin-Elbaum2004}
The pristine scrolls, depicted in Fig.~\ref{fig_dvomodel}(a), are
semiconductors with an optical gap of $\sim$0.5 eV. \cite{Cao2004}
This gap derives from a superposition of both V on-site $d$ to $d$
excitations and V$^{4+}$ to V$^{5+}$ charge transfer excitations.
\cite{Cao2004} Thus charge disproportionation exists already in the
pristine material. Optical gaps with similar magnitude are observed
in bulk VO$_2$, ladder-like $\alpha$$^\prime$-NaV$_2$O$_5$, tubular
Na$_2$V$_3$O$_8$, and certain molecular
magnets.\cite{Verleur1968,Long1999,Choi2002b,Barbour2006}
Within the rigid band scenario, the metal exchange process merely
involves charge injection into the V 3$d$ conduction band, which is
now partially-filled (see
Fig.~\ref{fig_dvomodel}(d)).\cite{Krusin-Elbaum2004}
The spectral response of the Mn$^{2+}$-substituted nanoscrolls in
Fig.~\ref{fig_dvolowfre}, however, clearly precludes the metallic
state expected from the bandfilling model.
Even if the partially-filled band of Fig.~\ref{fig_dvomodel}(d) is
split due to
chemical disorder effects (Fig.~\ref{fig_dvomodel}(e)), the
resulting low energy excitation would be fairly broad due to joint
density of states considerations in which the narrow occupied level
below $E_F$ is
coupled to the wider empty
conduction band arising from the V $3d$ states.

We believe that the optical response indicates that the ion exchange
process injects charges that are strongly localized due to disorder
and Madelung energy considerations. Any charge introduced into the
system can in principle reduce a V$^{5+}$ to a V$^{4+}$ ion. This
extra electron remains pinned, however, as Mn$^{2+}$ ions will
``bind'' preferentially (though not entirely) to the oxygen atoms
bonded to a V$^{4+}$ ion than to a V$^{5+}$ ion, since proximity to
a V$^{5+}$ ion would raise the Madelung energy substantially. Thus
even in the ```doped'' system there exists a barrier to electron
hopping between neighboring V$^{4+}$ and V$^{5+}$ ions.
Within the effective band model therefore the charging process leads
to strongly localized deep defect levels as shown in
Fig.~\ref{fig_dvomodel}(b). The d.c. conductivity may be dominated
by hopping between the disordered V$^{4+}$ and V$^{5+}$ ions.
The nonbonding character of the localized levels gives them their
narrow widths. The local potentials of the two kinds of ions
continue to be different due to electron-electron repulsions between
neighboring V$^{4+}$-V$^{4+}$ ions and long-range Madelung energy
involving  the Mn$^{2+}$ ions.
We have depicted this in Fig.~\ref{fig_dvomodel}(c) by an energy
splitting between ``occupied'' (predominantly V$^{4+}$) and
``unoccupied'' (predominantly V$^{5+}$) defect levels.
A dipole-allowed transition between these split defect levels would
be consistent with a localized charge
excitation rather than a traditional metallic response in the low
energy region of the optical conductivity. Such a feature would be
relatively narrow, consistent with the 400 - 500 cm$^{-1}$ linewidth
observed in the spectrum of Mn$^{2+}$ substituted scrolls.
The above localized picture is in better
accord with our experimental results for the pristine and Mn$^{2+}$
exchanged scrolls.

Careful examination of  vibrational 
mode patterns can also provide information on the ion exchange
process, although in this case, we must account for both charge and
structural modifications. This analysis is possible because
 vibrational features are  observed in both the pristine
and ion exchanged materials. In the pristine vanadium oxide
nanoscrolls, the  mode at $\sim$575 cm$^{-1}$ is assigned as the
V-O-V equatorial stretch. \cite{Cao2004} This mode is very sensitive
to curvature, widening as the size of the amine template is reduced,
characteristic of a more strained lattice. \cite{Cao2004} The V-O-V
equatorial stretching mode (Fig.~\ref{fig_dvolowfre}) redshifts with
ion substitution, moving from $\sim$575 cm$^{-1}$ in the pristine
scrolls, to $\sim$570 cm$^{-1}$ in Na$^+$ and $\sim$560 cm$^{-1}$ in
Mn$^{2+}$ exchanged compounds. Similar shifts in peak position are
also  observed in other doped and intercalated vanadates.
\cite{Cazzanelli1994, Choi2002}  
From the structural point of view, the redshifted V-O-V equitorial
streching mode in the scrolled vanadates is unexpected because x-ray
results indicate that the interlayer distance decreases as the
smaller metal ions replace the larger
amines. \cite{Reinoso2000} 
The V-O-V equatorial stretching mode might therefore be expected to
widen in the substituted scrolls due to their increased curvature,
but the center peak position ought to be relatively insensitive to
size. \cite{Cao2004} Ion exchange  also adds
charge to the scrolls. 
We attribute changes in the V-O-V equatorial stretching mode
position (Fig.~\ref{fig_dvolowfre}) to charging effects which
overcome the  aforementioned local strain that tends to widen and
slightly harden the V-O-V resonance. This overall softening trend
($\sim$575 cm$^{-1}$ in the pristine scrolls to $\sim$560 cm$^{-1}$
in the Mn$^{2+}$ substituted compound) confirms that the ion
exchange process adds carriers to the scrolls. The sharp, unscreened
vibrational features in the ion substituted scrolls show, however,
that the carriers are not mobile, consistent with the observation of
a
pinned charge excitation in the far infrared. The metal exchange
process also leads to some chemical disorder, as evidenced in the
splitting of a low frequency bending motion ($\sim$179 cm$^{-1}$)
and the screw-like mode  of the scroll at $\sim$112 cm$^{-1}$ that
is analogous to the radial breathing mode in carbon nanotubes
(inset, Fig.~\ref{fig_dvolowfre}). Disorder in Cu$^{2+}$ exchanged
nanoscrolls has been reported recently as well. \cite{Filho2004}

\begin{figure}[t]
\includegraphics[width=3.8 in]{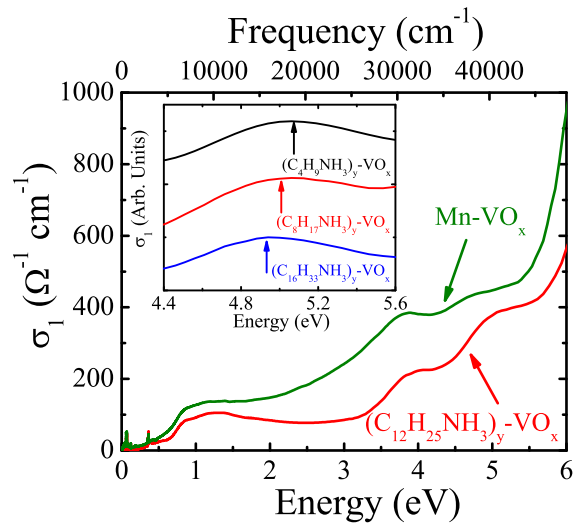}
\caption{\label{fig_dvoelectr}(color online) Expanded view of the
300 K optical conductivity of pristine VO$_x$ scrolls and the
Mn$^{2+}$ exchanged compound. The inset shows a close-up view of the
$\sim$5 eV excitation, which changes modestly with sheet distance in
the unsubstituted scrolls.\cite{Cao2004} In the inset, the curves
are offset for clarity. }
\end{figure}

Figure~\ref{fig_dvoelectr} displays the  300 K optical conductivity
of the pristine VO$_x$ scrolls and the Mn$^{2+}$ exchanged compound
over a broader energy range. With the exception of the
aforementioned bound carrier excitation localized in the far
infrared (Fig.~\ref{fig_dvolowfre}), the spectrum of the ion
exchanged scrolls is similar to that of the pristine compound, in
line with the fact that these excitations derive from the vanadium
oxide framework. \cite{Cao2004} Both materials display similar
$\sim$0.5 eV optical  gaps and excitations centered near 1 eV that
derive from a superposition of V $d$ to $d$ on-site excitations and
V$^{4+}$ to V$^{5+}$ charge transfer excitations.\cite{Cao2004} The
O $p$ to V $d$ charge transfer excitations at $\sim$3.9 and
$\sim$5.0 eV  redshift with ion substitution.  That this is a charge
(rather than size) effect is evident from a comparison of the data
in the inset of Fig.~\ref{fig_dvoelectr}. Here, the excitation
centered at $\sim$5.0 eV, which is  probably polarized in the radial
direction, shifts to higher energy as the size of the amine template
decreases (corresponding to a reduction in the interlayer
distance).\cite{Cao2004}
 The 3.9 eV feature does not change
with curvature. \cite{Cao2004} The Mn$^{2+}$ substituted scrolls
display the opposite trend, indicating that charge effects also
modify the charge transfer excitations. First principles electronic
calculations will offer microscopic insight on this problem.


We measured the  optical spectra of 
pristine and Mn$^{2+}$ substituted vanadium oxide nanoscrolls in
order to understand the  charge dynamics in the pristine and metal
exchanged materials and to test the applicability of the
rigid band model. 
In contrast to expectations from the rigid band model, the spectra
display a pinned low-energy electronic excitation in the Mn$^{2+}$
substituted nanoscrolls rather than a traditional metallic response.
We propose
that the injected charge is localized due to disorder and Madelung
energy effects. Our model is consistent with inhomogeneous charge
disproportionation (V$^{4+}$ and V$^{5+}$) in both the pristine and
the ion-exchanged materials, and explains the observed far infrared
charge localization.
%
Analysis of the vibrational properties shows that the 575 cm$^{-1}$
V-O-V equatorial stretching mode redshifts with ion substitution,
indicating that ion exchange modifies both the local curvature and
charge environment. Charge effects also redshift two high energy
electronic excitations.

\section*{Acknowledgements}

Work at the University of Tennessee is supported by the Materials
Science Division, Office of Basic Energy Sciences at the U.S.
Department of Energy under Grant DE-FG02-01ER45885. Research at
Binghamton is supported by the Division of Materials Research,
National Science Foundation under Grant 0313963. Work at University
of Arizona is supported by Materials Science Division, Office of
Basic Energy Sciences at the U.S. Department of Energy under Grant
DE-FG02-06ER46315. We thank H. Zhao for constructing Fig. 1(b).

\end{document}